\documentclass[%
aip,
amsmath,amssymb,
reprint,
]{revtex4-2}
\usepackage{caption}
\usepackage[table]{xcolor}
\usepackage{multirow}
\usepackage[toc,page]{appendix}
\usepackage{braket}
\usepackage{amsmath}
\usepackage{amsfonts}
\usepackage{amssymb}
\usepackage{graphicx}
\usepackage[colorlinks,linkcolor=blue,
anchorcolor=blue,citecolor=blue,
urlcolor=magenta]{hyperref}
\usepackage{mathrsfs}
\usepackage{dcolumn}
\usepackage{bm}
\usepackage{tensor}
\usepackage{color}
\usepackage{epsfig}
\usepackage{placeins}
\usepackage{textcomp}
\usepackage{bbold}
\usepackage{float}
\usepackage{verbatim}
\usepackage{times}
\usepackage{siunitx}
\usepackage{chemformula}
\usepackage{physics}
\usepackage{multirow}
\usepackage{makecell}
\usepackage{ragged2e}
\usepackage{subcaption}
\usepackage{gensymb}
\usepackage{makecell,booktabs}
\usepackage{color, colortbl}
\definecolor{mygray}{gray}{.9}
\definecolor{pink}{rgb}{.6902,.18824,.37647}
\usepackage{multirow}
\usepackage{array}
\usepackage{graphicx}
\usepackage{etoolbox}
\usepackage[utf8]{inputenc}
\usepackage[T1]{fontenc}
\usepackage[utf8]{inputenc}

\makeatletter
\def\@email#1#2{%
	\endgroup
	\patchcmd{\titleblock@produce}
	{\frontmatter@RRAPformat}
	{\frontmatter@RRAPformat{\produce@RRAP{*#1\href{mailto:#2}{#2}}}\frontmatter@RRAPformat}
	{}{}
}%
\makeatother

\renewcommand{\eqref}[1]{\textup{{\normalfont
			Eq.~(\ref{#1}}\normalfont)}}

\definecolor{c1}{HTML}{b71a3d}

\begin{document}
	
	\preprint{AIP/123-QED}
	
	\title[Quantum Advantage of One-Way Squeezing in Enhancing Weak-Force Sensing]{Quantum Advantage of One-Way Squeezing in Enhancing Weak-Force Sensing}

\author{Jie Wang}
\affiliation{Key Laboratory of Low-Dimensional Quantum Structures and Quantum Control of Ministry of Education, Department of Physics and Synergetic Innovation Center for Quantum Effects and Applications, Hunan Normal University, Changsha 410081, China}

\author{Qian Zhang}
\affiliation{Key Laboratory of Low-Dimensional Quantum Structures and Quantum Control of Ministry of Education, Department of Physics and Synergetic Innovation Center for Quantum Effects and Applications, Hunan Normal University, Changsha 410081, China}

\author{Ya-Feng Jiao}
\affiliation{Academy for Quantum Science and Technology, Zhengzhou University of Light Industry, Zhengzhou 450002, China}

\author{Sheng-Dian Zhang}
\affiliation{Key Laboratory of Low-Dimensional Quantum Structures and Quantum Control of Ministry of Education, Department of Physics and Synergetic Innovation Center for Quantum Effects and Applications, Hunan Normal University, Changsha 410081, China}

\author{Tian-Xiang Lu}
\affiliation{College of Physics and Electronic Information, Gannan Normal University, Ganzhou 341000, Jiangxi, China}

\author{Zhipeng Li}
\affiliation{Department of Electrical and
	Computer Engineering, National University of
	Singapore, Singapore 117583, Singapore}

\author{Cheng-Wei Qiu}
\affiliation{Department of Electrical and
	Computer Engineering, National University of
	Singapore, Singapore 117583, Singapore}

\author{Hui Jing}
\email{jinghui73@foxmail.com}
\affiliation{Key Laboratory of Low-Dimensional Quantum Structures and Quantum Control of Ministry of Education, Department of Physics and Synergetic Innovation Center for Quantum Effects and Applications, Hunan Normal University, Changsha 410081, China}

\date{\today}

\begin{abstract}
	Cavity optomechanical (COM) sensors, featuring efficient light-motion couplings, have been widely used for ultra-sensitive measurements of various physical quantities ranging from displacements to accelerations or weak forces. Previous works, however, have mainly focused on  reciprocal COM systems. Here, we propose how to further improve the performance of quantum COM sensors by breaking reciprocal symmetry in purely quantum regime. Specifically, we consider a spinning COM resonator and show that by selectively driving it in opposite directions, highly nonreciprocal optical squeezing can emerge, which in turn provides an efficient way to surpass the standard quantum limit that otherwise exists in conventional reciprocal devices.  Our work confirms that breaking reciprocal symmetry, already achieved in diverse systems well beyond spinning systems, can serve as a new strategy to further enhance the abilities of advanced quantum sensors, for applications ranging from testing fundamental physical laws to practical quantum metrology.
\end{abstract}
\maketitle

\begin{figure*}[htb]
	\center
	\includegraphics[width=1.95\columnwidth]{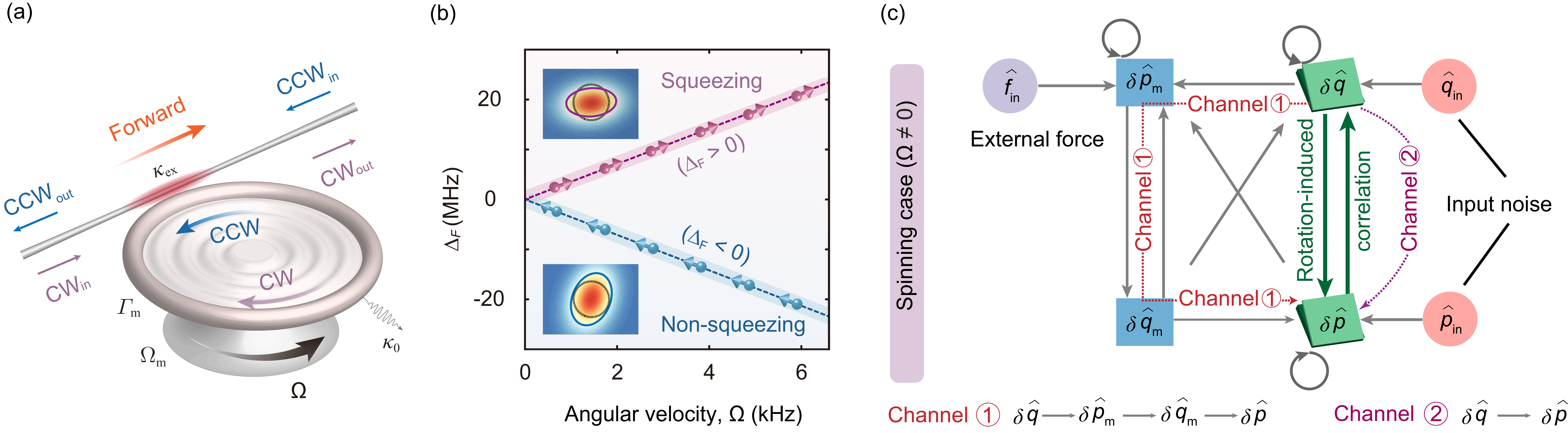}
	\caption{Enhancing quantum sensing by spinning a COM resonator. (a) The resonator supports two degenerate optical modes and a mechanical radial breathing mode. (b) The Sagnac-Fizeau shift $\Delta_{F}$ as a function of angular velocity $\Omega$. For a resonator spinning along the CCW direction, we have $\Delta_{F}>0$ or $\Delta_{F}<0$ for the forward (left) or backward (right) input direction, respectively. (c) A flow chart that describes the interaction relations in this system. For a static device, the optical quadratures $\delta\hat{q}$ and $\delta\hat{p}$ are indirectly coupled via the mechanical mode, corresponding to the Channel \textcircled{1}. The frequency shift $\Delta_{F}$ induces a direct coupling channel for $\delta\hat{q}$ and $\delta\hat{p}$, i.e., the Channel \textcircled{2}. By driving the device from one specific direction, it can lead to destructive interference between two channels, providing a way to suppress quantum noise below the SQL.}
	\label{Fig1}
\end{figure*}

\section{Introduction}

Quantum-enhanced sensing, featuring ultrahigh precisions enabled by unique quantum resources such as squeezing or entanglement, has been successfully demonstrated with diverse systems ranging from atoms or photons~\cite{pedrozo2020entanglement, mccormick2019quantum,wu2023quantum, chalopin2018quantum} to solid-state circuits~\cite{castellanos2008amplification, backes2021quantum, Xu2022Metrological} or mechanical devices~\cite{chen2023quantum, gilmore2021quantum}. In particular, rapid advances in cavity optomechanics (COM)~\cite{gavartin2012hybrid, tse2019quantum, yap2020broadband,clark2016observation, pirkkalainen2015squeezing} have provided unprecedented tools for detecting ultra-weak forces or tiny displacements~\cite{LiOuLeiLiu2021}. Notably, the standard quantum limit (SQL) can be reached for COM sensors, by balancing backaction noises and  shot noises (due to optical amplitude and phase fluctuations) \cite{kampel2017improving}. Such a SQL can be beaten by applying non-classical lights~\cite{xia2023entanglement, lawrie2019quantum}, imprecision-backaction correlations~\cite{ mason2019continuous,yu2020fihuantum,Sudhir2017} or using exquisite skills such as quantum non-demolition or backaction-evading measurements~\cite{mercier2021quantum,moller2017quantum,CavesEvading2012}. For a recent striking example, sub-SQL displacement sensing was demonstrated by injecting a quantum squeezed light into a macroscopic COM system even with a $\SI{40}{\kilo\gram}$ mirror~\cite{yu2020quantum}.

In parallel, nonreciprocal devices enabling one-way flow of photons have also attracted intense interests, due to their indispensable roles in applications ranging from backaction-immune communications to quantum measurements~\cite{shoji2014magneto}. Efficient magnetic-free strategies have been developed for building such devices at chip-scales with, e.g., spatiotemporal modulations~\cite{yu2009complete,kang2011reconfigurable, estep2014magnetic, tzuang2014non}, directional nonlinearities~\cite{dong2015brillouin, kim2015non, xia2018cavity, Rosario2018non,tang2022squeeze,Manipatruni2009optical, shen2016experimental,bernier2017nonreciprocal,de2019realization, yu2023integrated}, or non-Hermitian structures~\cite{peng2014parity, huang2021loss,fruchart2021non}. In a recent experiment, an optical diode with 99.6 $ \% $ isolation was demonstrated by spinning a resonator~\cite{SpinningResonator_Carmon}. Such spinning devices were then used to reveal directional quantum correlation effects such as nonreciprocal photon blockade or COM entanglement, with the merit of directionally enhanced robustness against random losses~\cite{huang2018nonreciprocal,Jiao2020Nonreciprocal}. Clearly, quantum correlation nonreciprocity is fundamentally different from classical mean-number nonreciprocity, since completely different correlations can emerge even for the same transmission rate. Hence, a classically reciprocal system can be highly nonreciprocal in non-classical level. In very recent experiments, nonreciprocal quantum correlations have been observed in cavity QED or solid-state systems~\cite{graf2022nonreciprocity,yang2023non}. However, as far as we know, the novel possibility of further enhancing quantum metrology by fully exploiting the merit of directional quantum correlations has not been explored to date.

Here, we propose to enhance quantum COM sensing by using nonreciprocal quantum squeezing. We show that by tuning the spinning velocity of COM resonator, the spinning-induced Sagnac effect can break the time-reversal symmetry of the COM system and mitigate the backaction noise induced by radiation pressure, resulting in highly asymmetric optical quadrature squeezing, i.e., the squeezing can emerge only in a chosen direction but not in the other direction. This provides a new degree of freedom to further suppress quantum noise below the SQL in a chosen direction, which is otherwise unattainable for its reciprocal counterpart. As a result, up to three orders of magnitude enhancement can be achieved, at least in principle, for the weak-force sensitivity under the same conditions, compared with that of a static device. Our work, showing the first example of the potential merit of nonreciprocal quantum-enhanced sensing, well compatible with state-of-the-art techniques of COM sensor fabrications, can inspire more efforts in the future  by combining more ideas of COM engineering, nonreciprocal physics, and quantum metrology.

\section{Results}\label{sec2}
\subsection{Spinning COM sensor}

In recent experiments, spinning devices have been used to realize directional heat flow~\cite{li2019anti,xu2020tunable}, resonator gyroscope~\cite{Mao:22}, and sound isolators~\cite{Sound2014Alu, Acoustic2019ding}. Particularly, an optical diode with 99.6 $ \% $ isolation was demonstrated by spinning an optical resonator~\cite{SpinningResonator_Carmon}, without relying on any magnetic materials or complex structures. Spinning systems have been used to predict nonreciprocal quantum correlation effects~\cite{huang2018nonreciprocal, Jiao2020Nonreciprocal}, which were then demonstrated with a solid-state device~\cite{graf2022nonreciprocity} and a cavity QED system~\cite{yang2023non} in very recent experiments. As a unique merit, nonreciprocal quantum correlations are more robust against random losses, in contrast to their reciprocal counterparts~\cite{huang2018nonreciprocal, Jiao2020Nonreciprocal}. However, the novel possibility of enhancing quantum metrology by utilizing such purely quantum nonreciprocal effects, as far as we know, has not been explored.

As shown in Fig.~\ref{Fig1}(a), we consider a COM resonator supporting two optical whispering-gallery-mode modes, i.e., the clockwise (CW) and counterclockwise (CCW) propagating modes, as well as a mechanical breathing mode. By spinning the resonator along a fixed direction at angular velocity $\Omega$, the resonance frequencies of the CW and CCW  modes experience opposite Sagnac-Fizeau shifts, i.e., $\omega_{0} \rightarrow \omega_{0}+\Delta_{F}$, with~\cite{malykin2000sagnac}	
\begin{equation}
	\Delta_{F}=\pm\frac{n
		R\Omega\omega_{0}}{c}\left(1-\frac{1}{n^{2}}
	-\frac{\lambda}{n}\frac{\mathrm{d}n}{\mathrm{d}\lambda}\right),
\end{equation}
where $\omega_{0}$ is the resonance frequency of the static resonator, $n$ is the refractive index of the material, $R$ is the resonator radius, and  $c$ ($\lambda$) is the speed (wavelength) of light in the vacuum. Usually, we can omit the dispersion term
$\mathrm{d}n/\mathrm{d}\lambda$
since it is relatively small in typical materials
($\sim 1\%$)~\cite{SpinningResonator_Carmon,malykin2000sagnac}. Moreover, as shown in Figs.~\ref{Fig1}(a) and \ref{Fig1}(b), by fixing the CCW rotation of the resonator, we have $\Delta_{F}>0$ ($\Delta_{F}<0$) for the situation where the driving field is input from the forward (backward) side, i.e.,  $\omega_{\mathrm{cw}, \mathrm{ccw}}=\omega_0 \pm\left|\Delta_F\right|$.

Experimental feasible parameters are chosen here~\cite{SpinningResonator_Carmon, righini2011Whispering}, i.e., a resonator with $Q=\num{3.2e7}$, corresponding to a total decay rate $\kappa/2\pi=\SI{6.43}{\mega\hertz}$, with both the decay rate $\kappa_{\mathrm{ex}}$  at the fiber-cavity interface and all other internal losses $\kappa_{\mathrm{0}}$,  and the cavity coupling efficiency $\eta_{c}=\kappa_{\mathrm{ex}}/(\kappa_{0}+\kappa_{\mathrm{ex}})$~\cite{Sudhir2017}; the effective mass of the mechanical mode is $m=\SI{10}{\nano \gram}$, with $Q_{m}=\num{1.21e4}$ and the associated damping rate $\Gamma_{m}=\SI{5.2}{\kilo\hertz}$. In a frame rotating at driving frequency $\omega_{l}$, quantum Langevin equations (QLEs) of this system can be written at the simplest level as
\begin{align}
	\dot{\hat{q}}_{m}
	= &~\Omega_{m}\hat{p}
	_{m},\nonumber \\
	\dot{\hat{p}}_{_{\mathrm{m}}} = &-\Omega_{m}\hat{q}_{m}
	-\Gamma_{m}\hat{p}_{m} -g_{0}\hat{a}^{\dagger} \hat{a}+\sqrt{2\Gamma_{m}}\hat{f}_{\mathrm{in}},\nonumber\\
	\dot{\hat{a}} = &
	-\left[i\left(\Delta_{c}+\Delta_{F}\right)
	+\frac{\kappa}{2}\right]\hat{a}-ig_{{0}}\hat{a}
	\hat{q}_{m}+{\cal E}\nonumber \\ &+\sqrt{\eta_{c}\kappa}\hat{a}_{\mathrm{in}}+\sqrt{\left(1-\eta_{c}\right)\kappa}\hat{a}_{0}.\label{QLEs}
\end{align}
where $\hat{a}$ ($\hat{a}^{\dagger}$) is the optical annihilation (creation) operator, $\hat{q}_{m}$ and $\hat{p}_{m}$ are the dimensionless position and momentum operators of the mechanical mode, respectively. $\Delta_{c}=\omega_{\mathrm{0}}-\omega_{l}$ is the detuning between the cavity and the driving field, $\Omega_{m}$ is the mechanical frequency, and $g_{0}$ is the COM coupling strength. The field amplitude of the driving laser is ${\cal E}$. As in the experiment~\cite{SpinningResonator_Carmon}, optical backscattering losses are excluded here. In fact, recent studies have confirmed that even strong backscattering losses can be directionally suppressed in spinning devices~\cite{Jiao2020Nonreciprocal, xiangSwitching2023}. Also, $\hat{a}_{\mathrm{0}}$ or $\hat{a}_{\mathrm{in}}$ describes the Gaussian noise from the vacuum or the cavity-fiber coupling port, respectively, and $\hat{f}_{\mathrm{in}}=\hat{f}_{\mathrm{th}}+\hat{f}_{\mathrm{sig}}$, with $\hat{f}_{\mathrm{th}}$ or $\hat{f}_{\mathrm{sig}}$ corresponding to the thermal Brownian force and the detected force signal, respectively. The input noise operators have zero mean values and the correlation functions
$$
\left\langle\hat{a}_{0}(t)\hat{a}_{0}^{\dagger}(t^{\prime})\right\rangle=\delta(t-t^{\prime}),~~ \left\langle\hat{a}_{\mathrm{in}}(t)\hat{a}_{\mathrm{in}}^{\dagger}(t^{\prime})\right\rangle=\delta(t-t^{\prime}),
$$
also, under thermal equilibrium, we have~\cite{bowen2015quantum}
$$
\left\langle\hat{f}_{\mathrm{th}}(t)\hat{f}_{\mathrm{th}}(t^{\prime})\right\rangle\simeq\bar{n}_{m}\delta(t-t^{\prime}),~~~ \text{for}~~ \bar{n}_{m},~ Q_{m}\!\gg\!1,$$
where $\bar{n}_{m}\approx k_{\mathrm{B}}T/\hbar \Omega_m$ and $k_{\mathrm{B}} $ or $T$ is the Boltzmann constant or the bath temperature~\cite{coherentheurs2014}, respectively.

In order to solve the QLEs, we expand each operator as the sum of its steady-state value and a small fluctuation around it under the strong-driving condition, i.e.,
$\hat{a}=\alpha+\delta \hat{a}, \hat{q}_{m}=\bar{q}_{m}+\delta\hat{q}_{m}, \hat{p}_{m}=\bar{p}_{m}+\delta\hat{p}_{m}$. Such a standrad procedure can lead to the steady-state mean values given by
\begin{align}
	|\alpha|^2 &=4\eta_{c}\kappa P_{\mathrm{in}}/[\hbar\omega_{l}(\kappa^{2}+4\tilde{\Delta}^{2})],\nonumber\\
	\bar{q}_{m} &=-g_{{0}}|\alpha|^2/\Omega_{m},~~ \bar{p}_{m}=0,
\end{align}
where $\tilde{\Delta}=(\Delta_{c}+g_{0}\bar{q}_{m})+\Delta_{F}$. The quadrature operators of optical fluctuations are defined as
$\delta\hat{q}=\frac{1}{\sqrt{2}}\left( \delta\hat{a}^{\dagger}+ \delta\hat{a}\right),
\delta\hat{p}=\frac{i}{\sqrt{2}}\left(\delta\hat{a}^{\dagger}-\delta\hat{a}\right)$,
with the associated noises
$\hat{q}^{\text {in, 0}}$ and $\hat{p}^{\mathrm{in, 0}}$. The resulting linearized QLEs can be written as
\begin{align}
	\label{noise-eq}
	\delta\dot{\hat{q}} = &-\frac{\kappa}{2}\delta\hat{q}+ \tilde{\Delta}\delta\hat{p}+g\sin\phi\delta\hat{q}_{m}+\sqrt{\eta_{c}\kappa}\delta
	\hat{q}^{\text {in}}\nonumber \\
	&+\sqrt{\left(1-\eta_{c}\right)\kappa}
	\delta\hat{q}^{\text {0}},\nonumber \\
	\delta\dot{\hat{p}} = &-\frac{\kappa}{2}\delta\hat{p}- \tilde{\Delta}\delta\hat{q}-g\cos\phi\delta\hat{q}_{m}+\sqrt{\eta_{c}\kappa}\delta
	\hat{p}^{\text {in}}\nonumber \\
	&+\sqrt{\left(1-\eta_{c}\right)\kappa}\delta\hat{p}^{\text {0}},\nonumber \\
	\delta\dot{\hat{q}}_{m} = &\Omega_{m}
	\delta\hat{p}_{m},\nonumber \\
	\delta\dot{\hat{p}}_{m} = &-\Omega_{m}\delta\hat{q}_{m}-\Gamma_{m}\delta\hat{p}_{m}
	-g\cos\phi\delta\hat{q}-g\sin\phi\delta\hat{p}\nonumber \\
	&+\sqrt{2\Gamma_{m}}\hat{f}_{\mathrm{in}},
\end{align}
where $g=\sqrt{2} g_0  \lvert \alpha\rvert$ and the cavity-field phase is chosen as $\phi=\arctan(-2\tilde{\Delta}/\kappa)$~\cite{bowen2015quantum}. The interaction relations underlying Eq.\,(\ref{noise-eq}) can be seen in a flow chart as plotted in Fig.\,\ref{Fig1}(c), where arrows connect the corresponding variables on the two sides of the linearized QLEs. To focus on the nonreciprocal effect induced by the Sagnac effect, we set $\Delta_{c}=-g_{0}\bar{q}_{m}$ and consider the case  $\tilde{\Delta}=\Delta_{F}$~\cite{SpinningResonator_Carmon,huang2018nonreciprocal,jiao2022non,xiangSwitching2023}.  For a static device, optical quadratures are indirectly coupled through the mechanical mode (the Channel \textcircled{1}, see Fig.\,\ref{Fig1}(c))~\cite{CavesCoherent2010}.  In contrast, a direct coupling between them can emerge due to the spinning-induced Sagnac effect (i.e., the Channel \textcircled{2}). The destructive interference of such two channels thus provides a way to suppress quantum backaction noise.

\begin{figure}[t!]
	\includegraphics[width=0.95\columnwidth]{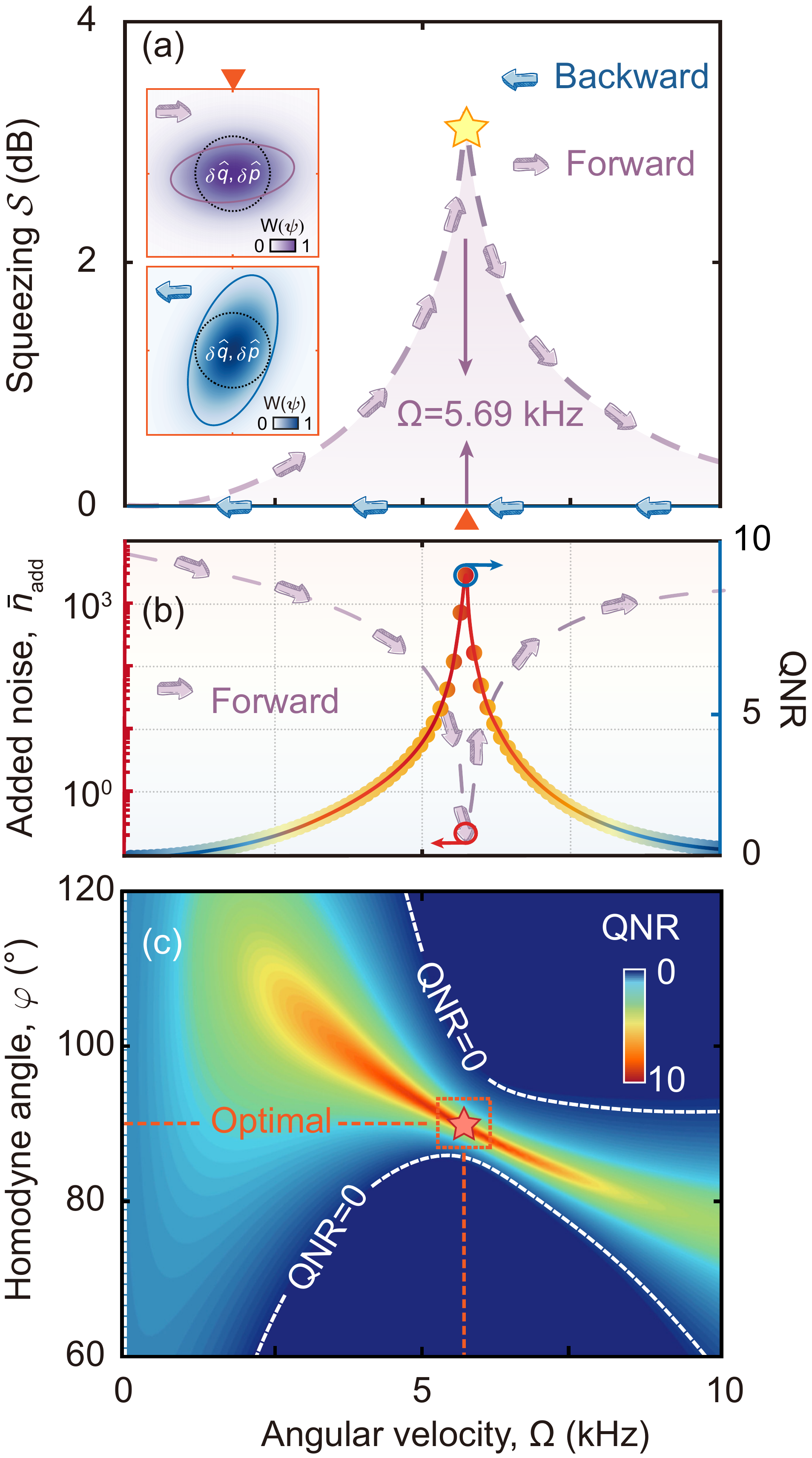}
	\caption{Suppressing quantum noise with nonreciprocal optical squeezing. (a) The degree of squeezing $\mathcal{S}$ is plotted as a function of angular velocity $\Omega$ for different input directions. The inserts  show the characteristic projections of $W(\psi)$ on two-dimensional subspace $\left\{\delta \hat{q}, \delta \hat{p}\right\}$ with $\Omega=\SI{5.69}{\kilo\hertz}$. The ellipse (circle) with the solid (dashed) line indicates a drop by 1/e of the maximum value of $W(\psi)$ for the steady (vacuum) state. (b) The added noise $\bar{n}_{\mathrm{add}}$ and quantum nonreciprocity ratio (QNR) as functions of $\Omega$, showing the lowest added noise at the optimal QNR. (c) QNR as a function of $\Omega$ and the homodyne angle $\varphi$ with the Fourier frequency $\omega/2\pi=\SI{1}{\kilo\hertz}$. The red star marks the position of the optimal QNR.}
	\label{figure2}
\end{figure}

\begin{figure*}[htb]
	\center
	\includegraphics[width=1.95\columnwidth]{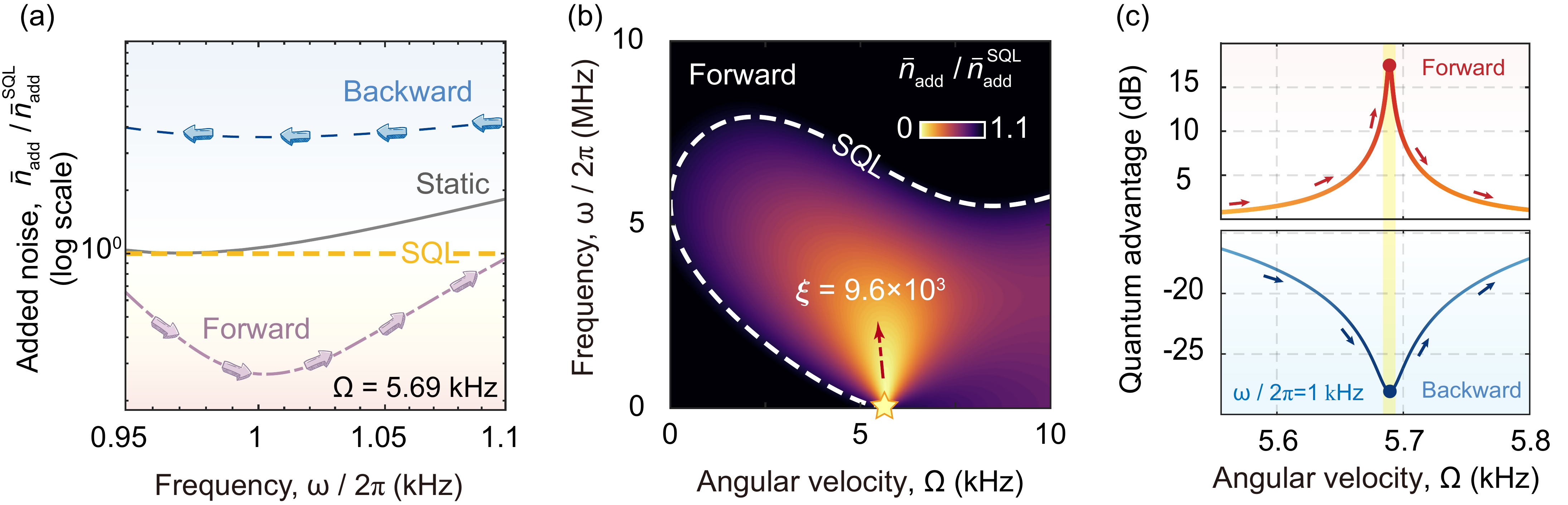}
	\caption{Achieving sub-SQL quantum sensing with nonreciprocal optical squeezing. (a) The scaled noise spectrum $\bar{n}_{\mathrm{add}}/\bar{n}_{\mathrm{add}}^{\mathrm{SQL}}$ in the frequency domain is plotted for different input directions, with $\Omega=0$ or $\Omega\neq 0$. The added noise $\bar{n}_{\mathrm{add}}$ can be suppressed well below the SQL for the forward input case, in the price of noise amplification for the backward input case. (b) $\bar{n}_{\mathrm{add}}/\bar{n}_{\mathrm{add}}^{\mathrm{SQL}}$ as a function of $\Omega$ and the Fourier frequency $\omega$ for the forward input case.  (c) Quantum advantage, as defined in the main text,  is plotted as a function of $\Omega$ for different input directions, also reaching optimal for the optimal QNR.}
	\label{figure3}
\end{figure*}

With the help of the input-output relation $\hat a_{\mathrm{out}}=\sqrt{\eta_c\kappa}\hat a-\hat a_{\mathrm{in}}$~\cite{Aspelmeyer2014cavity}, we can readout the force noise via homodyne detection~\cite{Sudhir2017, Clerk2010Introduction,Sudhir2017}. Under this procedure, the transmitted field is mixed with a local oscillator with a phase $\varphi$ at a $50$:$50$ beam splitter.
The photocurrent $\hat{I}_{\varphi}$ at the output of the balanced homodyne detector is proportional to a rotated field quadrature, $\delta\hat{q}_{\mathrm{out}}^{\varphi}[\omega]=\delta\hat{q}_{\mathrm{out}}\cos\varphi+\delta\hat{p}_{\mathrm{out}}\sin\varphi$.  	The output amplitude quadrature spectrum of the system  in
frequency space is defined as 
\begin{equation}
	\bar{S}_{\mathrm{qq}}^{\mathrm{out}}\left[\omega\right]=\frac{1}{2}\left\langle \left\{ \delta\hat{q}_{\mathrm{out}}\left[\omega\right],\delta\hat{q}_{\mathrm{out}}
	\left[\omega^{\prime}\right]\right\} \right\rangle,
\end{equation}
and similar forms hold for $\bar{S}_{\mathrm{pp}}^{\mathrm{out}}$ and $\bar{S}_{\mathrm{pq}}^{\mathrm{out}}$~\cite{Sudhir2017}. The homodyne spectrum of the output field is then derived as~\cite{Sudhir2017}: $\bar{S}_{\mathrm{II}}^{\varphi}[\omega] \propto \frac{1}{2}\left\langle\left\{\delta \hat{q}_{\mathrm{out}}^{\varphi}[\omega], \delta \hat{q}_{\mathrm{out}}^{\varphi}[\omega^{\prime}]\right\}\right\rangle=\mathcal{R}_m[\omega]\left(\bar{n}_m+\bar{n}_{\mathrm{add}}[\omega]\right),$
where $\mathcal{R}_m\qty[\omega]
=\abs\big{\partial\hat{q}_{\mathrm{out}}
	^{\varphi}\qty[\omega]\big/\partial\hat{f}
	_{\mathrm{sig}}\qty[\omega]}^{2}$ is the mechanical response of the sensor to the detected force signal. The added noise, including both the shot noise and the backaction noise, contributes to the total force noise spectrum quantifying the force sensitivity, i.e.,
\begin{equation}
	\bar{S}_{\mathrm{FF}}[\omega]=2\hbar m
	\Gamma_{m}\Omega_{m}(\bar{n}_{m}+\bar{n}_{\mathrm{add}}). \label{sff}
\end{equation}
Thermal noise contributions are subtracted to reveal the influence of the Sagnac effect on the quantum noise and the performance of the spinning COM sensor.

\subsection{Nonreciprocal quantum-enhanced force sensing}

The quantum squeezing can be observed in the output spectrum, by choosing an appropriate homodyne angle $\varphi$~\cite{Purdysqueezing2013}, as given by the output-field squeezing spectrum
\begin{align}
	\mathcal{S}_{qz}& =\int \mathrm{d} \omega^{\prime}\left\langle\hat{q}_{\mathrm{out}}^{\varphi}(\omega) \hat{q}_{\mathrm{out}}^{\varphi}\left(\omega^{\prime}\right)\right\rangle,
\end{align}
and the degree of squeezing, quantified in decibels ($\si{\decibel}$), is
\begin{align}\label{squeezing}
	\mathcal{S}=-\log _{10}2\mathcal{S}_{qz}\left[\omega\right].
\end{align}
The output field is squeezed for $\mathcal{S}_{qz}<1/2$ or equivalently $\mathcal{S}>0$.
As Fig.\,\ref{figure2}(a) shows,  when the output of a static device has no squeezing ($\mathcal{S}=0$), one-way squeezing can occur for the spinning case. Specifically, when driving the system in the forward direction, nonreciprocal squeezing can be achieved in such a  spinning device. This effect of directional squeezing effect can be intuitively understood by examining the Wigner function~\cite{Adesso2004extremal}
\begin{equation}
	W(\psi)=\frac{\exp[-\frac{1}{2}(\psi
		\mathcal{V}^{-1}\psi^{\dagger})]}{\pi^{2}\sqrt{
			\operatorname{det}\mathcal{V}}},
\end{equation}
with $\psi=(\delta \hat{q},\,\delta  \hat{p},\,
\delta \hat{q}_{m},\,\delta \hat{p}_{m})$, the state
vector of the quadrature fluctuations.  $\mathcal{V}$ can be  determined by the  Lyapunov equation~\cite{Vitali2007entanglement}. The inserts of Fig.\,\ref{figure2}(a) shows the projections of the reconstructed $W(\psi)$ for local quadrature pairs $\left\{\delta \hat{q}, \delta \hat{p}\right\}$. We see that for the backward input case, such a projection is characterized by a typical Gaussian distribution while for the forward input case, non-classical quadrature squeezing can emerge for the optical mode, which is consistent with the nonreciprocal signature of $\mathcal{S}$ as revealed above. To clearly see the link between nonreciprocal squeezing and added noise suppression, we can define the quantum nonreciprocity ratio 
\begin{equation}
	\text{QNR}=-\log _{10}\frac{2\mathcal{S}_{qz}\left[\omega\right](\Delta_{F}>0)}
	{2\mathcal{S}_{qz}\left[\omega\right](\Delta_{F}<0)},
\end{equation}
and QNR $=0$ corresponds to the reciprocal case. In Fig.\,\ref{figure2}(b), we plot the added noise $\bar{n}_{\textrm{add}}$ and the QNR as a function of $\Omega$ with the same parameters. We see that optimal values of $\bar{n}_{\textrm{add}}$ and the QNR are achieved at $\Omega=5.69\,\textrm{kHz}$ and the increase of the QNR leads to significant suppression of the added noise. The optimal position of the QNR is also located in Fig.\,\ref{figure2}(c), a plot of the QNR as a function of $\Omega$ and $\varphi$.

Now we show how to improve the weak-force sensitivity by exploiting the directional optical squeezing. As Fig.\,\ref{figure3} shows, for the standard COM sensing procedure (i.e., $\eta_{c}=1$, $\varphi$= $\pi$/2, and $\Omega=0$), the force sensitivity in Eq.\,(\ref{sff}) is constrained by the SQL, as indicated by the symmetrized noise spectrum (for $\kappa\gg\omega$)
\begin{align}
	\bar{n}_{\mathrm{add}}&=\bar{n}_{\mathrm{add}}^{\mathrm{shot}}+\bar{n}_{\mathrm{add}}^{\mathrm{qba}}=\frac{g^{2}}{\kappa \Gamma_{m}}+\frac{1}{16} \frac{\kappa}{g^{2} \Gamma_{m}} \frac{1}{\left|\chi_{m}\right|^{2}},\label{nadd}
\end{align}
with the mechanical susceptibility $\chi_{m}=\Omega_{m}/(\Omega_{m}^{2}
-\omega^{2}-i\omega\Gamma_{m})$. 
This equation shows a tradeoff between shot noise and backaction noise in the force sensing procedure, and the minimum added noise can be derived as
\begin{equation}
	\bar{n}_{\mathrm{add}}^{\mathrm{SQL}}[\omega]=\frac{1}{2
		\Gamma_{m}\abs{\chi_{m}}},
\end{equation}
which is known as the SQL~\cite{Aspelmeyer2014cavity}.  
As shown in Fig.\,\ref{figure3}(a), the scaled noise spectrum $\bar{n}_{\mathrm{add}}/\bar{n}_{\mathrm{add}}^{\mathrm{SQL}}$ is plotted for different input directions. We see that with the same parameters, $\bar{n}_{\mathrm{add}}$ is above the SQL for the static or reciprocal case, while it can be directionally suppressed well below the SQL for the spinning or nonreciprocal case. Such noise suppression thus can lead to enhanced force sensitivity [see Eq.\,(\ref{sff})]. To find the optimal sensitivity, we plot $\bar{n}_{\mathrm{add}}/\bar{n}_{\mathrm{add}}^{\mathrm{SQL}}$ as a function of $\Omega$ and the Fourier frequency $\omega$ in Fig.\,\ref{figure3}(b), showing the minimum added noise at $\omega/2\pi=\SI{1}{\kilo\hertz}$, $\Omega=\SI{5.69}{\kilo\hertz}$. Defining the sensitivity enhancement factor as
\begin{equation}
	\xi=\frac{\min \left\{\bar{S}_{\mathrm{FF}}[\omega](\Omega=0)\right\}}{\min \left\{\bar{S}_{\mathrm{FF}}[\omega](\Omega \neq 0)\right\}},
\end{equation}
we find that it can be enhanced by three orders of magnitude, compared with the static case. Alternatively, we can also define the quantum advantage of our spinning system over its static counterpart as the following ratio
\begin{equation}
	\text{quantum advantage}\,(\text{dB})=10\log _{10} \sqrt{\frac{\bar{S}_{\mathrm{FF}}^{\mathrm{SQL}}[\omega](\Omega=0)}
		{\bar{S}_{\mathrm{FF}}[\omega](\Omega \neq 0)}},
\end{equation}
where $\bar{S}_{\mathrm{FF}}^{\mathrm{SQL}}[\omega](\Omega=0)$ denotes the best sensitivity of the static COM sensor~\cite{li2021quantum}. Figure \ref{figure3}(c) shows this ratio as a function of $\Omega$ for different input directions. Clearly, quantum advantage exists only for the forward case, reaching an optimal value over $15\,\textrm{dB}$. This result coincides well the occurrence of directional optical squeezing [see Fig.\,\ref{figure2}], facilitated by the destructive two-channel interference of the noises (see Fig.\,\ref{Fig1}(c)].

\section{Conclusion}

In summary, we have proposed to enhance quantum noise suppression by utilizing nonreciprocal quantum squeezing, with the specific example of quantum weak-force sensing. We show that the spinning-induced Sagnac effect can lead to not only breaking of time-reversal symmetry of the COM system, but also destructive two-channel interference of quantum noise. As a result, directional optical squeezing can be realized in such a nonreciprocal quantum sensor, enabling sub-SQL quantum noise suppression and three-order enhancement of force sensitivity in a chosen direction, which is otherwise unattainable for the corresponding static or reciprocal system. Further improvement of our system is also possible by e.g., incorporating intracavity squeezing medium or injecting an external squeezed light into the COM sensor as already demonstrated in experiments~\cite{squeezing2015Lyu,Exponentially2018Qin,Peano2015intracavity,zhao2020weak}. Our work not only provides a new strategy to improve quantum-enhanced metrology by using different types of nonreciprocal devices, such as levitated COM devices~\cite{kuang2023nonlinear} and nonreciprocal atomic or solid-state systems~\cite{xiacavity2018, yang2023non, luatoms2021,yu2020experimental, zhang2019quantum, wang2023non, Ding2023limit, graf2022nonreciprocity}, but also builds a bridge between such diverse fields as COM engineering, nonreciprocal physics, and quantum metrology.

\section*{Supplementary Material}

we present more details in the supplementary material of quantum advantage of one-way squeezing in enhancing weak-force sensing, including: (1)  detailed derivations of the linearized Hamiltonian (section S1);
(2) noise spectrum (section S2); (3) more discussions on experimental feasibility (section S3); (4) reconstructed Wigner function (section S4); (5) thermal noise (section S5).

\section*{Acknowledgements}

Hui Jing is supported by the National Natural Science Foundation of China (NSFC, Grants No. 11935006 and No. 11774086), the Science and Technology Innovation Program of Hunan Province (Grant No. 2020RC4047), and the Hunan Provincial Major Sci-Tech Program (Grant No. 2023ZJ1010). Ya-Feng Jiao is supported by the National Natural Science Foundation of China (NSFC, Grant No. 12147156). Tian-Xiang Lu is supported by the National Natural Science Foundation of China (NSFC, Grant No. 12205054), and Ph.D. Research Foundation (BSJJ202122). Cheng-Wei Qiu acknowledges the support from National Research Foundation (award number: NRF2021-QEP2-03-P09 $\&$ WBS number: A-8000708-00-00).

\section*{Author Declarations}	

\subsection*{Conflict of Interest}
	
The authors have no conflicts to disclose.
	
	\subsection*{Author Contributions}

\textbf{Jie Wang}: Data Curation (equal), Formal Analysis (equal), Methodology (equal), Software (equal), Validation (equal), Visualization (equal), Writing/Original Draft Preparation (equal), Writing/Review \& Editing (equal). 
\textbf{Qian Zhang}: Data Curation (equal), Formal Analysis (equal), Methodology (equal), Software (equal), Validation (equal), Visualization (equal), Writing/Review \& Editing (equal).
\textbf{Ya-Feng Jiao}: Conceptualization (support), Methodology (support), Software (support), Writing/Original Draft Preparation (equal), Writing/Review \& Editing (equal).
\textbf{Sheng-Dian Zhang}:  Data curation (support), Methodology (support), Software (support).
\textbf{Tian-Xiang Lu}:  Data curation (support), Conceptualization (support), Methodology (support), Software (support).
\textbf{Zhipeng Li}:  Data curation (support), Methodology (support), Writing/Original Draft Preparation (support), Writing/Review \& Editing (support).
\textbf{Cheng-Wei Qiu}: Supervision(support), Validation (support), Writing/Original Draft Preparation (support), Writing/Review \& Editing (support).
\textbf{Hui Jing}: Conceptualization (lead), Methodology (support), Resources (equal), Supervision (equal), Validation (equal), Funding Acquisition (equal),  Writing/Original Draft Preparation (equal), Writing/Review \& Editing (equal). 

\section*{Data Availability Statement}
The data that support the findings of this study are available from the corresponding author upon reasonable request.

\section*{References}
\end{document}